\documentclass{article}
\usepackage[preprint]{neurips_2021}
\usepackage[utf8]{inputenc} % allow utf-8 input
\usepackage[T1]{fontenc}    % use 8-bit T1 fonts
\usepackage{hyperref}       % hyperlinks
\usepackage{url}            % simple URL typesetting
\usepackage{booktabs}       % professional-quality tables
\usepackage{amsfonts}       % blackboard math symbols
\usepackage{amsthm}
\usepackage{nicefrac}       % compact symbols for 1/2, etc.
\usepackage{microtype}      % microtypography
\usepackage{xcolor}         % colors
\usepackage{dirtytalk}
\usepackage{physics}
\usepackage{graphicx}
\usepackage{todonotes}
\usepackage{svg}

\newtheorem{definition}{Definition}

\title{Partial sensitivity analysis in differential privacy
}

\author{Tamara T. Müller \dag\ddag, Alexander Ziller \dag\ddag, Dmitrii Usynin \dag\ddag\S, Moritz Knolle \dag\ddag\S,\\\textbf{Friederike Jungmann \dag\ddag, Daniel Rueckert \ddag\S, Georgios Kaissis \ddag\S\P}\\\ddag Technical University of Munich, \S Imperial College London,\\ \dag equal contribution, \P \url{g.kaissis@tum.de}}

\begin{document}

\maketitle

\begin{abstract}
Differential privacy (DP) allows the quantification of privacy loss when the data of individuals is subjected to algorithmic processing such as machine learning, as well as the provision of objective privacy guarantees.
However, while techniques such as individual Rényi DP (RDP) allow for granular, per-person privacy accounting, few works have investigated the impact of each input feature on the individual's privacy loss.
Here we extend the view of individual RDP by introducing a new concept we call \textit{partial sensitivity}, which leverages symbolic automatic differentiation to determine the influence of each input feature on the gradient norm of a function.
We experimentally evaluate our approach on queries over private databases, where we obtain a feature-level contribution of private attributes to the DP guarantee of individuals. Furthermore, we explore our findings in the context of neural network training on synthetic data by investigating the partial sensitivity of input pixels on an image classification task.
\end{abstract}

\section{Introduction}
The handling and processing of sensitive data inherently entails the risk of compromising individual privacy and exposing personal information. 
The differential privacy (DP) framework \cite{dwork2014algorithmic} allows to reason about privacy loss in the setting of data processing and provide privacy guarantees to individuals while allowing to draw conclusions from the dataset as a whole.
However, certain attributes of the individual which serve as input variables, may contribute more to their identifiability in a dataset than others. Moreover, the individuals themselves may perceive some of their attributes as being more sensitive and thus more worthy of protection from exposure. Transparent and trustworthy data processing systems must therefore be capable of not only accounting for individual privacy loss (e.g. shown in work on individual Rényi DP (RDP) \cite{feldman2020individual}), but to also selectively apply privacy-preserving mechanisms to specific private attributes \cite{alaggan2015heterogeneous}. So far, few works have investigated the contribution of individual attributes to the query function's sensitivity, and therefore to overall privacy loss. This is, in part, because in existent machine learning frameworks, it is impossible to disentangle the interaction between input features and model components such as weights. To address this issue, recent work has introduced symbolic automatic differentiation (AD)-based sensitivity analysis \cite{ziller2021sensitivity}.
The main contribution of our work is the introduction of partial sensitivity, representing the fractional contribution of each input feature to the gradient norm, which determines privacy loss in DP, and can be derived using symbolic AD. We demonstrate this technique in the contexts of queries over private databases and neural network training on synthetic data. Moreover, we showcase the markedly different influence of non-private stochastic gradient descent (SGD) on partial sensitivity compared to DP-SGD. 

\section{Related Work}
\label{sec:prior_work}

Typically, DP guarantees are allotted homogeneously across individuals in the dataset, only considering the worst case scenario (i.e. where privacy accounting is carried out using global sensitivity). As, however, the realised gradient norm of a query may be considerably lower, the outlook of this type of privacy accounting may be unnecessarily pessimistic. Individual privacy accounting, and specifically the work by Feldman et al. \cite{feldman2020individual}, proposes an alternative formulation which separately accounts for each individual's influence on the outcome of a computation via the actual gradient norm. The authors express privacy guarantees using Rényi DP \cite{mironov2017renyi}. We note that the utilisation of the gradient norm for individual privacy accounting, which we will be using in this work, is performed in conjunction with a privacy filter or odometer presented in \cite{feldman2020individual}. Heterogeneous DP \cite{alaggan2015heterogeneous} captures the variation of privacy expectations across information held by one individual and uses a modified Laplace mechanism. Phan et al. \cite{Phan2019, Phan2017} address the distribution of noise between different input variables depending on their importance to the network output using the Laplace or the Heterogeneous Gaussian Mechanism. Our work is complementary to these techniques as partial sensitivity captures the influence of individual input attributes on the gradient norm and can therefore be employed to guide noise addition to selectively perturb them using one of these heterogeneous noise mechanisms.

\section{Background}
\label{sec:definitions}
We briefly introduce key terminology, and assume familiarity with the concepts of differential privacy, the Gaussian mechanism \cite{dwork2014algorithmic} and (DP-)SGD \cite{abadi2016deep}.
We will use the following notations: $D$ and $D'$ for two adjacent datasets, whereby $\sim$ denotes add/remove one adjacency. We will use $f$ for a function/query and $\Delta_2(f)$ for its $L_2$-sensitivity.

\begin{definition}[$L_2$-sensitivity]
The $L_2$-sensitivity of a function $f$ is defined as the maximum change in $L_2$-norm in the image of $f$ over all pairs of adjacent inputs:
\begin{equation}
    \Delta_2(f) = \max_{D, D': D \sim D'}  \Vert f(D) - f(D') \Vert_2
    \label{def:sensitivity}
\end{equation}
\end{definition}

The $L_2$-sensitivity is used to calibrate the noise addition in the Gaussian Mechanism. If $f$ is Lipschitz continuous, its Lipschitz constant is equivalent to its sensitivity \cite{raskhodnikova2016lipschitz}:

\begin{definition}[Lipschitz continuity]
Let $f$ be a function $X \rightarrow Y$ with associated metrics $d_X$ and $d_Y$. $f$ is Lipschitz continuous if and only if there exists a real constant $K \geq 0$ such that, $\forall x_1, x_2 \in X$:
\begin{equation}
    d_Y(f(x_1),f(x_2)) \le K d_X(x_1,x_2)
\end{equation}
The Lipschitz constant of a real-valued function is equal to the supremum of the norm of its gradient: $K(f) = \sup_x \Vert\nabla f(x) \Vert_2$.
\label{def:lip}
\end{definition}

\section{Theoretical Results}
\label{sec:methods}

\begin{definition}[\textbf{Partial sensitivity of a function}]
Let $f: \mathbb{R}^m \rightarrow \mathbb{R}^1, m>1$ be a real-valued function. The partial sensitivity of $f$ at an input $X$ is defined as the gradient of the gradient norm of $f$ with respect to the input variables:
\begin{equation}
    \Delta_p (f(X)) := \nabla_{X} \Vert \nabla f(X) \Vert_2
    \label{def:partial_sensitivity}
\end{equation}
\end{definition}

The definition of partial sensitivity is derived from the following property:
Let $\vb{x} = \left(x_1, x_2, \dots, x_d \right)^T$ be the vector of inputs. Since $f: \mathbb{R}^m \rightarrow \mathbb{R}^1$, $f(\vb{x})$ is a scalar. As such, $\nabla f(\vb{x}) = \left (\frac{\partial f(\vb{x})}{\partial x_1} , \dots, \frac{\partial f(\vb{x})}{\partial x_d} \right )$. Furthermore, let $\left (\frac{\partial f(\vb{x})}{\partial x_i} \right )=a_i, i\leq d$ and let  $\varphi$ be the $L_2$-norm of  $\nabla{f(\vb{x})}$, given by $\varphi = \Vert \nabla{f(\vb{x})} \Vert_2 = \sqrt{\sum_{i=1}^{d} a_i^2}$ which is also a scalar. As we are interested in the rate of change of the sensitivity with respect to the individual input attributes, $(x_1, \dots, x_d)$, we obtain the gradient of $\varphi$:
\begin{align}
    \nabla \varphi = \left (\frac{\partial}{\partial x_1} \sqrt{\sum_{i=1}^{d} a_i^2}, \dots, \frac{\partial}{\partial x_d} \sqrt{\sum_{i=1}^{d} a_i^2} \right )
\end{align}
The $j_{th}$ component of $\nabla\varphi$, $\nabla^j\varphi$, where $j\leq d$ is given by:
\begin{align}
    \frac{\partial }{\partial x_j} \sqrt{\sum_{i=1}^{d} a_i^2} = 
    \frac{1}{2}\frac{2a_j}{\sqrt{\sum_{i=1}^{d} a_i^2}} = 
    \frac{a_j}{\varphi} = 
    \frac{a_j}{\Vert \nabla{f(\vb{x})} \Vert_2} =
    \frac{\frac{\partial f(\vb{x})}{\partial x_j}}{\Vert \nabla{f(\vb{x})} \Vert_2} 
\end{align}
Thus:
\begin{align}
    \Delta^{(x_1, \dots, x_d)}_p(f(\vb{x})) =
    \left ( \frac{\frac{\partial f(\vb{x})}{\partial x_1}}{\Vert \nabla{f(\vb{x})} \Vert_2},  
    \dots,
    \frac{\frac{\partial f(\vb{x})}{\partial x_d}}{\Vert \nabla{f(\vb{x})} \Vert_2} \right )
\end{align}
Partial sensitivity therefore represents the fractional contribution of the individual input attributes to the gradient norm of the function. Its symbolic representation, which can be easily obtained using a symbolic AD system \cite{ziller2021sensitivity}, is independent of the actual input data and can therefore be used to interpret the impact of individual input attributes on the gradient norm, used for (individual) privacy accounting, as shown in the next section.

\section{Experiments}
\label{sec:experiments}
\subsection{Partial sensitivity analysis of multivariate database queries}
\label{sec:priv_query}
\begin{figure}
    \centering
    \includegraphics[scale=0.5]{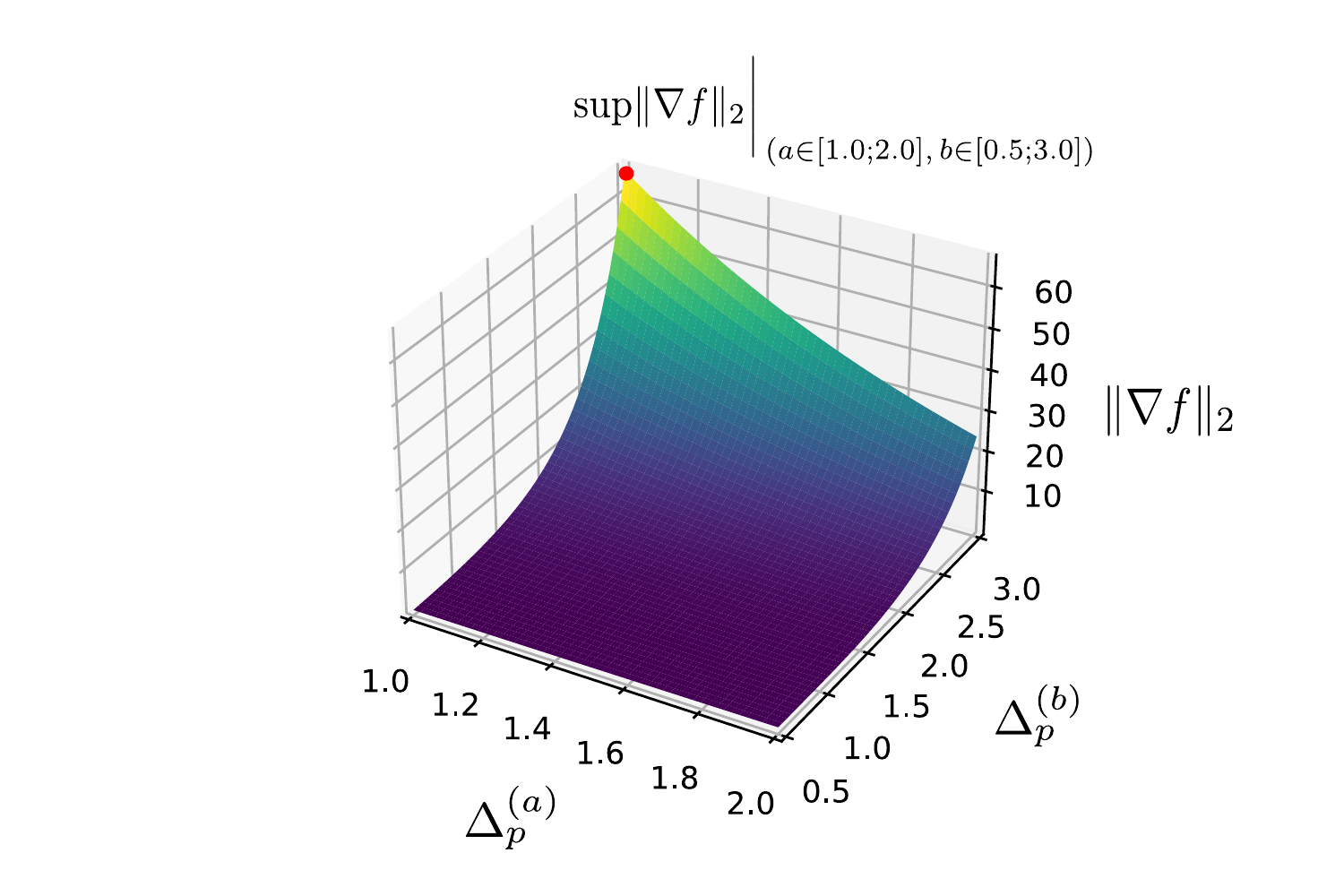}
    \caption{Partial sensitivity plot for section \ref{sec:priv_query}. The horizontal axes show the partial sensitivity of $f$ with respect to inputs $a$ and $b$. The vertical axis shows $\Vert  \nabla f_i \Vert_2$. $\Delta_2(f)$ given the input ranges is marked in red.}
    \label{fig:3d_plots}
\end{figure}

We assume a scenario where an analyst wants to construct a statistical query $f(a,b)$ over a database containing private data from a population. They specify reasonable ranges on the private attributes $a \in [1.0, 2.0], b \in [0.5, 3]$ (e.g. based on prior knowledge but not on the actual data values). Let $f(a, b) = a^2 + e^{2 b - a}$ and $\phi: X \rightarrow \mathbb{R}^1$ be a linear aggregation function with sensitivity $1$. To obtain the Rényi DP guarantee for one individual $i$ as specified in Section \ref{sec:prior_work}, the Lipschitz constant $(L_{\phi})_i$ of $\phi$ and the norm of the gradient of $f$ of each individual $i$, $\Vert \nabla_i f\Vert_2$, are required. Following Definition 2.7 from \cite{feldman2020individual} and Definition \ref{def:lip}, the linearity of $\phi$ means that the same Lipschitz constant $L_{\phi} = \sup \Vert \nabla \phi \Vert_2 = 1$ can be used for all individuals.
The privacy loss of individual $X_i$ can then be calculated as follows: $(\alpha, \frac{\alpha L_{\phi}^2 \Vert \nabla f_i \Vert_2^2}{2\sigma^2}) = (\alpha, \frac{\alpha \Vert  \nabla f_i \Vert_2^2}{2\sigma^2})$, where $\alpha$ is the \textit{Rényi} divergence order and $\sigma^2$ the variance of the Gaussian noise added by the mechanism. Furthermore, compiling the symbolic expression for the gradient norm of $f$ and maximising its value using a suitable technique such as \textit{simplicial homology global optimisation} \cite{endres2018simplicial}, the global sensitivity $\Delta_2(f)$ can be calculated given the pre-defined ranges of the input variables $a$ and $b$. This has to be done data-independently, and $\Delta_2(f)$ is a constant (in this example $\Delta_2(f) = 66.19$). Finally, the partial sensitivity of $f$, $\Delta^{(a, b)}_p(f)$, can be analysed and plotted to permit visual reasoning about the effects of $a$ and $b$ on $\Vert \nabla_i f\Vert_2$, and thus on Rényi DP guarantees. Figure \ref{fig:3d_plots} shows how the two variables contribute to the sensitivity of $f$. This knowledge allows to assign a different weight to each attribute and calibrate noise addition to selectively protect specific attributes.

\subsection{Partial sensitivity modelling in (DP-)SGD}
\begin{figure}
    \centering
    \includegraphics[width=0.8\textwidth]{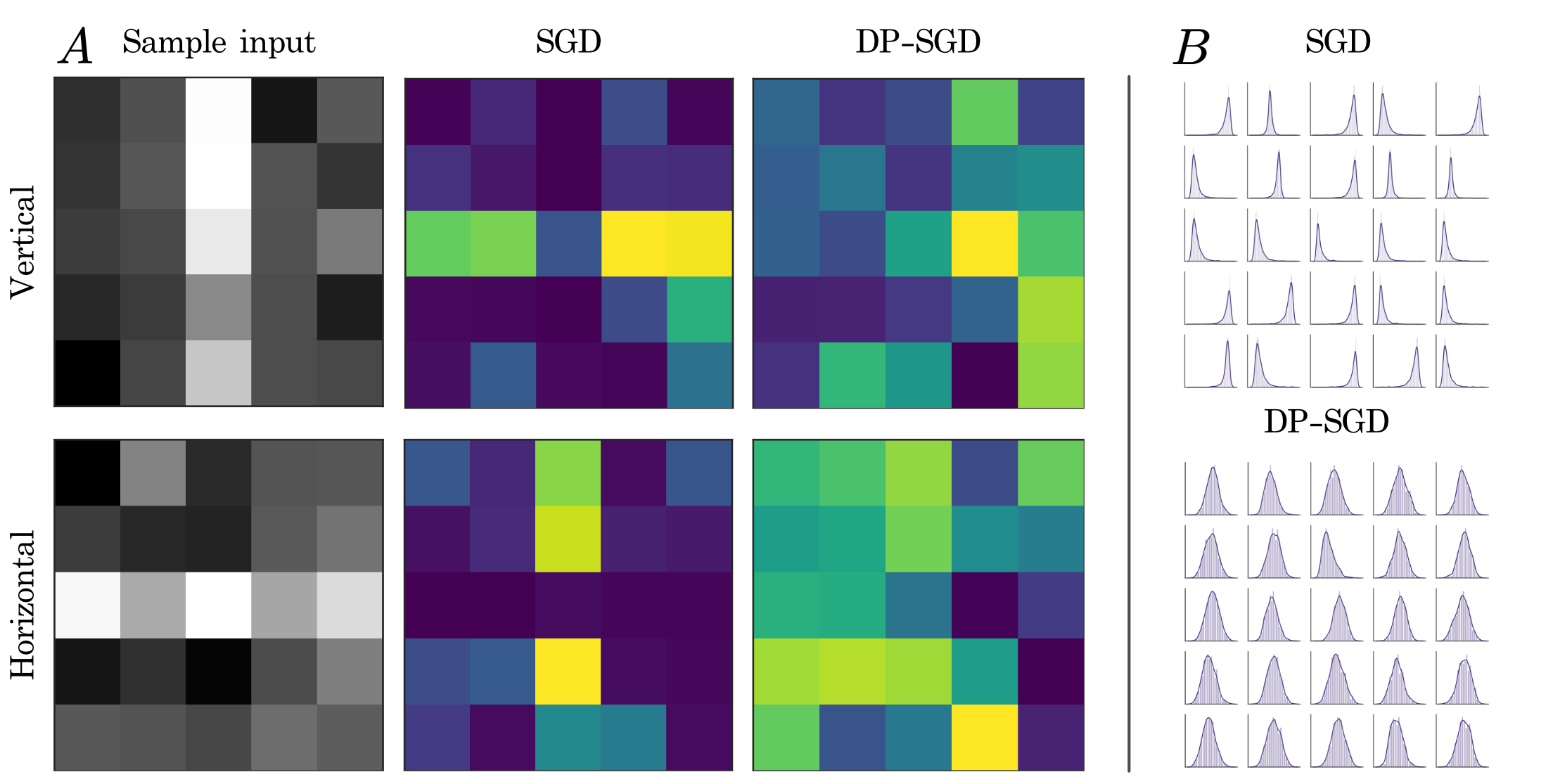}
        \caption{Pixel-wise privacy loss analysis of neural networks. \textbf{A} shows input images of the vertical and horizontal bar class, the maximum partial sensitivities for each pixel for SGD and DP-SGD. \textbf{B} shows histograms of the partial sensitivity of each pixel using SGD and DP-SGD.}
    \label{fig:pixelwise_sens}
\end{figure}

In this section, we investigate the impact of individual input attributes on the privacy loss of a specific neural network training example. To illustrate this effect, we performed binary classification on synthetic $5\times5$-pixel images of vertical and horizontal bars using SGD and DP-SGD using a two-layer neural network. The symbolic representation for the partial sensitivity of the loss function, $\Delta_p^{(x_{1,1}, \dots, x_{5,5})}(\mathcal{L})$ can again be derived through symbolic AD. By substitution of the weights, class label and input pixel values, it is possible to obtain the partial sensitivity of the input pixels at any moment in training. At the end of training, where we consider the weights to be fixed, this can be used to reason over the relevant input features leading the network to assign an image to one of the two classes. For the vertical bar class, the maximum partial sensitivity values were observed across the horizontal of the image, and vice versa for the horizontal bar class (Figure \ref{fig:pixelwise_sens}\textbf{A}). Evidently, the presence of features characteristic for the opposite class renders the individual an \say{outlier} in the specific distribution, resulting in a large change in the function's gradient norm and thus, a high individual privacy loss. Interestingly, training with DP-SGD led to a substantial reduction in the partial sensitivities of the corresponding pixels, while increasing the values of unrelated pixels in the input space. To further illustrate this effect, we generated histograms of the partial sensitivities for each pixel over one thousand samples. We found the distributions of the partial sensitivities to be highly concentrated around specific values in the case of SGD, while being substantially more dispersed and centered around zero for DP-SGD (Figure \ref{fig:pixelwise_sens}\textbf{B}). From this we conjecture that the effect of DP-SGD training is the \say{homogenisation} of the partial sensitivity across the space of inputs, which may suppress the memorisation of strongly identifying private attributes. \footnote{Source code at: \url{https://github.com/tamaramueller/Deuterium_Partial_Sensitivity}}

\section{Discussion and Conclusions}
\label{sec:discussion}
Our work extends sensitivity analysis to the level of individual input attributes by introducing \textit{partial sensitivity}. Individuals may wish to weight their specific attributes in a personalised manner as they consider them to have a different privacy impact, or even allocate fractions of their privacy budget to different analysts based on such feature-level granularity. Analysts on the other hand, may be interested to attenuate the specific features leading to high privacy expenditure, as demonstrated in our case study on DP-SGD. Our results may help to better understand the privacy-utility trade-offs of this algorithm \cite{dp_impact_accuracy, dp_bad_2, avent2019automatic}, its impact on fairness \cite{fioretto2021decision}, and the link between memorisation of private features, model overfitting, and generalisation \cite{feldman2020does}. Moreover, partial sensitivity exhibits strong links to gradient-based model introspection techniques \cite{gradcam}.  We intend to explore these topics, and the utilisation of partial sensitivity to guide targeted noise addition via heterogeneous mechanisms, in future work.

\bibliographystyle{unsrt}
\bibliography{bibliography}
\newpage

\appendix

\section{Appendix}

\paragraph{Description of computational resources used}
All experimentation was conducted on a single workstation computer with 18 physical CPU cores and 512 GB of physical memory, as well as 2 NVidia Quadro RTX 8000 GPUs running \textit{Ubuntu Linux} v. 20.04 LTS. 

\paragraph{Model architectures}
For the partial sensitivity modelling in DP-SGD, we designed a feedforward neural network consisting of an input layer of dimensionality $25 \times 8$, an intermediate layer of dimensionality $8 \times 8$ with a bias vector of length $8$ and an output layer of dimensionality $8 \times 1$. All layers were followed by \textit{logistic sigmoid} activation functions. Training was performed using the binary cross-entropy loss function.

\paragraph{Synthetic data generation}
Synthetic data generation was conducted as follows: \newline
Two base images with white background were constructed, where one consisted of a vertical bar of black pixels and the other of a horizontal line. From these, $1000$ images per class were created by the addition of random Gaussian noise with mean zero and a standard deviation of $0.2$. Evaluation was performed on a test set of $100$ images per class. A fixed random seed was used to deterministically generate the images in the experiments of the main manuscript.

\paragraph{Hyperparameters}
We trained all networks to convergence using SGD with a batch size of $2000$ (\textit{batch gradient descent}) and a learning rate $\eta=0.1$. We repeated this procedure using DP-SGD with the same learning rate, a noise multiplier of $5.0$ and an $L_2$-bound of $0.1$.

\paragraph{Details on the attached code}
Symbolic automatic differentiation was performed using the \textit{Deuterium} framework \cite{ziller2021sensitivity}, whose source code is available alongside the experiment code. Compilation performed by \textit{Deuterium} relies on a suitable compiler for the \textit{C} or \textit{Fortran} programming language or for the \textit{LLVM} tool-chain.

\end{document}